\documentclass[runningheads]{llncs}

\usepackage{graphicx}
\usepackage{multirow} 
\usepackage{pifont}
\usepackage{hyperref}
\usepackage{booktabs}
\usepackage{array}
\usepackage{amssymb}
\usepackage{enumitem} 
\usepackage{amsmath}
\usepackage{subcaption}
\usepackage{tabularx}
\usepackage{graphicx}
\usepackage{booktabs,multirow,tabularx,array}
\usepackage{booktabs,tabularx,array,adjustbox}
\newcolumntype{Y}{>{\raggedright\arraybackslash}X}

\usepackage{xcolor}

\begin{document}
\title{A Multi-Agent Approach to Validate and Refine LLM-Generated Personalized Math Problems}

\author{
Fareya Ikram\inst{1}\orcidID{0000-0003-0444-6870}\and
Nischal Ashok Kumar\inst{1}\orcidID{0000-0002-7022-5948} \and
Junyang Lu\inst{1}\orcidID{0009-0007-3258-6447} \and
Hunter McNichols\inst{1}\orcidID{0009-0002-0129-260X} \and
Candace Walkington\inst{2}\orcidID{0000-0002-2338-8760} \and
Neil Heffernan\inst{3}\orcidID{0000-0002-6185-5841}\and
Andrew S.\ Lan\inst{1}\orcidID{0000-0002-8475-6600}
}

\authorrunning{F. Ikram et al.}

\institute{
University of Massachusetts Amherst, Amherst, MA, USA
\and
Southern Methodist University, Dallas, TX, USA
\and
Worcester Polytechnic Institute, Worcester, MA, USA \\
\email{fikram@umass.edu}
}

\titlerunning{A Multi-Agent Approach for Personalized Math Problems}
\maketitle              

\begin{abstract}
Students benefit from math problems contextualized to their interests. Large language models (LLMs) offer promise for efficient personalization at scale. However, LLM-generated personalized problems may often have problems such as unrealistic quantities and contexts, poor readability, limited authenticity with respect to students' experiences, and occasional mathematical inconsistencies. To alleviate these problems, we propose a multi-agent framework that formalizes personalization as an iterative generate--validate--revise process; we use four specialized validator agents targeting the criteria of solvability, realism, readability, and authenticity, respectively. We evaluate our framework on 600 problems drawn from a popular online mathematics homework platform, ASSISTments, personalizing each problem to a fixed set of 20 student interest topics. We compare three refinement strategies that differ in how validation feedback is coordinated into revisions. Results show that authenticity and realism are the most frequent failure modes in initial LLM-personalized problems, but that a single refinement iteration substantially reduces these failures. We further find that different refinement strategies have different strengths on different criteria. We also assess validator reliability via human evaluation. Results show that reliability is highest on realism and lowest on authenticity, highlighting the need for better evaluation protocols that consider teachers' and students' personal characteristics.
\end{abstract}

\section{Introduction}

Context personalization, tailoring instructional content to align with students’ individual interests, has been shown to increase engagement and learning outcomes across diverse educational settings \cite{lin2024personalized}. This task has long been constrained by the effort required for teachers and curriculum developers. The rise of large language models (LLMs) offers potential for educational tasks to be personalized to fine-grained student interests at scale \cite{bahel2025personalizing}. However, research cautions that personalization will not effectively facilitate learning unless it connects to students' authentic real-world experiences \cite{walkington2020appraising}.

Current LLM-based personalization tools face several technical challenges that may limit their adoption in classroom settings. Studies of teacher interactions with these systems identify four recurring challenges. First, personalized problems may contain mathematical inaccuracies, such as invalid solutions, inconsistent quantities, or ambiguous setups \cite{christ-etal-2024-mathwell}\cite{einarsson2024}. Second, these problems may include unrealistic quantities or scenarios that violate real-world norms \cite{walkington2025efficiency} \cite{walkington2025middle}. Third, the problems often lack authenticity, failing to align with students' actual interests or lived experiences \cite{walkington2025efficiency}. Finally, research suggests LLM-generated content may increase reading burdens on students \cite{christ2025edumath} \cite{oh2026classroom}. 

In order to address the challenge of balancing multiple constraints, researchers have explored iterative refinement and structured control of agents \cite{Jeong2025}. Iterative refinement strategies like Self-Refine \cite{madaan2023self} and Reflexion \cite{shinn2023reflexion} demonstrate that LLMs can improve outputs through critique-and-revise loops, relying on undifferentiated self-feedback that may overlook domain-specific constraints. 

We address the multi-faceted nature of context personalization by proposing a multi-agent approach that employs specialized validation and refinement agents for personalizing math problems. Recent work in the field of generative AI has demonstrated that multi-component agentic LLM pipelines outperform single-prompt approaches on complex generation tasks with multi-dimensional criteria \cite{Barrak2025}. However, a key question remains on how validation feedback, especially for the task of context personalization, should be coordinated into effective revisions. For example, two criteria, such as adding realistic contextual details while maintaining appropriate readability, need to be balanced. In this work, we experiment with a multi-agent framework that employs iterative generate-validate-revise cycles with four specialized validator agents, i.e., agents targeting each of the identified challenges. We compare three refinement strategies: (1) Centralized Refinement, which aggregates feedback from all validator agents before revision; (2) Centralized Refinement with Planning, which adds a planning component to structure and prioritize feedback; and (3) Decentralized Refinement, which couples each validator agent with a dedicated refinement agent. 

We evaluate our framework by first comparing the performance of three different refinement strategies. Second, we assess the reliability of our validator agents via human evaluation. Our results show that realism and authenticity are the most common failure modes in initial LLM-personalized problems, but that a single refinement iteration substantially reduces these errors. We further find that refinement behavior is criterion-dependent: Decentralized Refinement reduces realism and readability failures more quickly, while Centralized Refinement with Planning converges fastest on authenticity. Finally, our human evaluation shows that the realism validator agent matches human judgments more reliably than the authenticity validator agent, suggesting that authenticity assessments often depend on an individual's familiarity with the topic. Our results motivate future evaluations grounded in teacher and student judgments.

\section{Related Works}
\subsubsection{Obstacles in LLM Generated Educational Content} Research on LLM-based educational personalization has documented specific patterns of concern. Several K-12 teacher-facing generative AI products, including Khanmigo \cite{khanmigo} and MagicSchool \cite{magicschool}, provide tools for adapting lessons to students' interests, yet teachers report these adaptations as insufficient. Research in mathematics education finds that typical mathematical tasks are often framed as contrived and unrealistic with respect to students' interests, and inauthentic with respect to students' actual experiences with quantities and measurement \cite{Palm2008}.

Studies of teacher-LLM interactions reveal that teachers frequently prompt systems to adjust unrealistic problem formulations where numbers violate real-world norms, for example, a bottle of water costing 100 dollars \cite{walkington2025efficiency} \cite{walkington2025middle}. Teachers also attend closely to popular culture references, noting when references like Fortnite resonate more with their students than alternatives like Roblox, demonstrating concern for contextual authenticity and timeliness. Additionally, teachers adjust readability levels of LLM output to meet their students' needs \cite{walkington2025efficiency}. While mathematical accuracy has improved since earlier investigation, contemporary models still generate problems with mathematical errors \cite{christ-etal-2024-mathwell}. 


\subsubsection{Multi-Agent Architectures} Multi-agent systems have emerged as a promising paradigm for educational content generation, demonstrating that complex educational tasks benefit from specialized, coordinated agents rather than single models. For instance, EduAgentQG \cite{jia-etal-2025-eduagentqg} demonstrates that multi-agent workflows can generate diverse, high-quality problems aligned to educational objectives through specialized agents for planning, writing, and validation. While such approaches advance question generation, creating new problems from scratch to cover learning objectives, our work addresses a fundamentally different and complementary challenge: context personalization. Rather than generating new problems, we adapt existing, validated problems to student interests while preserving their underlying mathematical structure and pedagogical intent.

Similarly, recent work on multi-agent collaborative frameworks for math problem generation \cite{karbasi2025multi} employs teacher-critic cycles where a teacher agent generates question-answer pairs and a generic critic provides iterative feedback on clarity, relevance, and difficulty alignment. Their evaluation reveals incremental improvements over baseline approaches, with the greatest gains observed in difficulty matching and relevance, highlighting the value of multi-agent refinement for educational content generation. However, their approach employs a single generic critic to assess all quality dimensions at once, which may limit the system's ability to provide targeted, criterion-specific feedback to guide revisions. 

The value of multi-agent approaches extends beyond question generation to other educational content tasks requiring multi-dimensional quality assessment. Yang et al. \cite{yang2024content} demonstrate this in teacher professional development, where their LLMAgent-CK framework employs multiple agents in collaborative discussion to assess teachers' mathematical content knowledge. Their multi-agent design addresses challenges that parallel those in personalization: response diversity, limited training data, and the need for interpretable assessments across multiple criteria. By distributing evaluation across specialized agents, they achieve more reliable judgments than single-model approaches, a principle that motivates our use of dedicated validators for distinct quality dimensions.

Specialized agent architectures reflect broader recognition that complex educational tasks benefit from explicit validation. Ensuring pedagogical standards requires explicit validation mechanisms \cite{scarlatos2024improving}. While prior work establishes the value of specialized validators, a key design question remains of how feedback from multiple validators should be coordinated into effective revisions.

\subsubsection{Iterative Refinement}
In order to address the challenge of balancing multiple constraints—from maintaining mathematical context while applying personalization, to restricting linguistic complexity while preserving realism—researchers have explored iterative refinement and structured control of agents. Researchers have proposed an adaptive refinement system where agents target specific aspects, such as factuality and personalization, and then merge them into a final, optimized response \cite{Jeong2025}. Techniques utilizing agentic context engineering enable systems that adapt their strategies over time, guarding against the potential loss of domain-specific constraints during iterative rewriting \cite{zhang-etal-2025-agentic-context}. Iterative refinement strategies like Self-Refine \cite{madaan2023self} and Reflexion\cite{shinn2023reflexion} demonstrate that LLMs can improve outputs through critique-and-revise loops, though these approaches typically rely on self-feedback that may overlook domain-specific constraints.

While prior work demonstrates the value of multi-agent systems for educational content generation, a gap remains in understanding how to systematically validate and refine personalized adaptations. Our work addresses this gap by focusing on context personalization for mathematics problems, where we identify four distinct quality criteria derived from teacher feedback and systematically compare alternative strategies for turning validation feedback into revisions.

\section{Multi-Agent Math Problem Personalization}\label{sec:method}
\begin{figure}[t]
    \centering
    \includegraphics[width=1\linewidth]{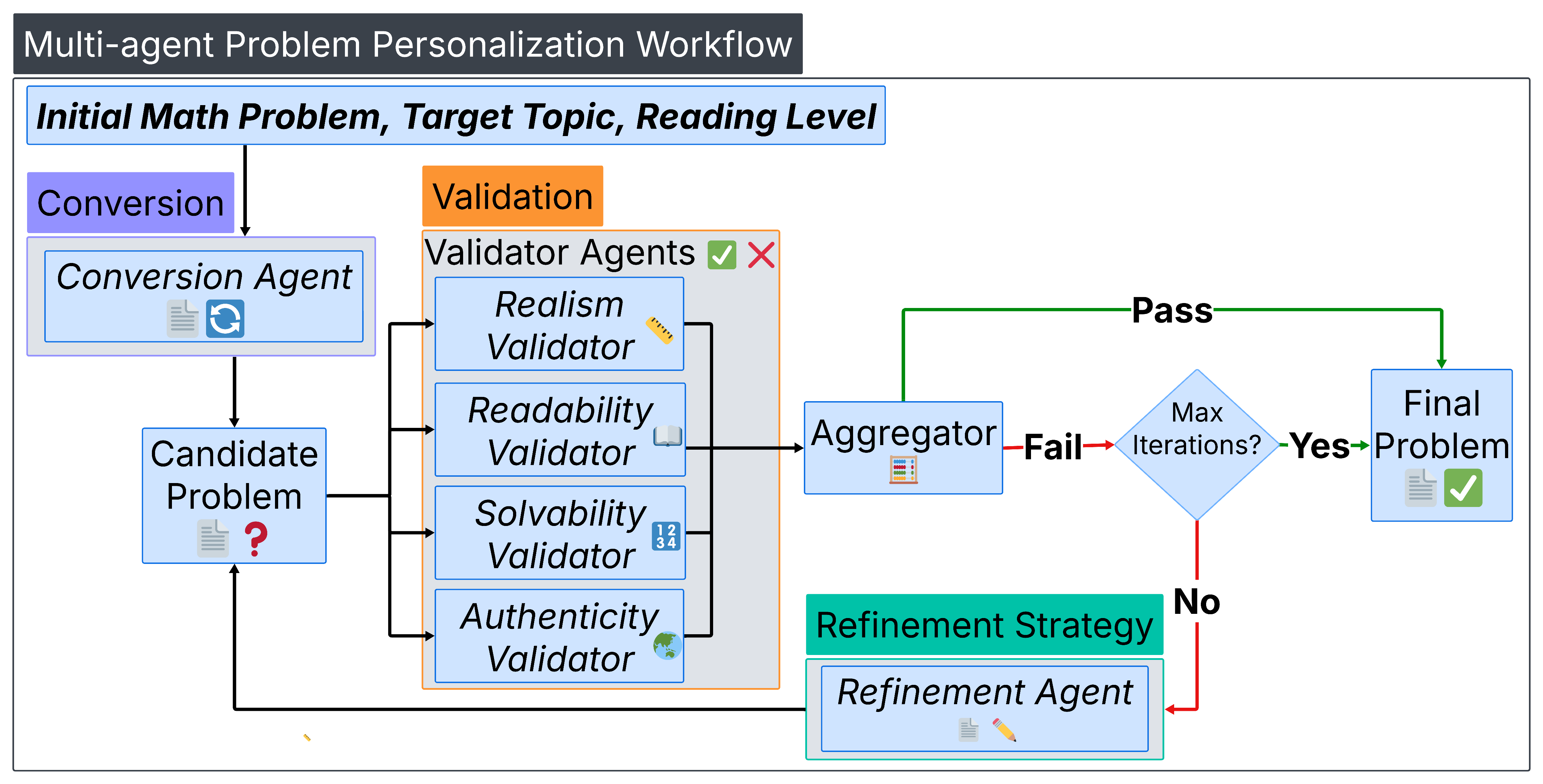}
    \caption{An overview of our multi-agent math problem personalization workflow utilizing the \textit{Centralized Refinement} strategy. Given an initial math problem, target topic, and reading level we utilize the \textit{Conversion Agent} to generate a candidate problem. The candidate problem is checked by four \textit{Validator Agents}. If all pass, we have the final problem. Otherwise, the refinement agent generates a new candidate problem given the validator agent feedback. This process repeats until the candidate problem passes all agents or meets the maximum number of refinement iterations.}
    \label{fig:placeholder}
\end{figure}

In this section, we define our math problem context personalization task and detail the corresponding multi-agent workflow.

\subsection{Task Definition}\label{sec:method:problem}
We study the task of personalizing a math problem to a target topic that reflects a learner's interest, while preserving the underlying mathematical structure and solution and grade-appropriate language and context.
Given an original problem $p$ and a target topic $t$, our system produces a revised candidate problem $\hat{p}$ via an iterative generate--validate--revise loop. 

\subsection{Workflow Overview} 

The framework consists of a \textit{conversion agent} that generates an initial topic-aligned candidate problem $\hat{p}$, four independent \textit{validator agents} that return a binary decision (\textit{pass/fail}) and brief diagnostic feedback $f$ describing the detected issues in the candidate $\hat{p}$, and one or more \textit{refinement agents} that revise the candidate problem using validator feedback. We detail these agents below. 

\subsubsection{Conversion Agent}\label{conv_agent}
The conversion agent takes in an original problem $p$, a target topic $t$, and a target grade reading level $g$ computed using the Flesch--Kincaid grade-reading level metric \cite{norberg2025linguistic}. This agent utilizes a prompt aimed to generate a new candidate $\hat{p}$ that satisfies four constraints: being realistic, authentic, matching the original reading level of problem $p$, and being mathematically consistent with the original math problem $p$. We refer to the output of this conversion agent that has not been passed through the validators as \textit{Zero-Shot}. Below, we detail the validators that target each of the four constraints. 

\subsubsection{Validator Agents}\label{sec:val_agents}
We now detail the four validator agents, aimed at reflecting common challenges faced during the context personalization of math problems.
\begin{itemize}
    \item \textit{Realism Agent}. The realism validator agent checks whether quantities, units, and contextual details mentioned in the candidate problem $\hat{p}$ are plausible within the target topic $t$(e.g., ensuring sports scores and time intervals fall within realistic ranges).
    \item \textit{Readability Agent}. The readability validator agent checks whether vocabulary, sentence complexity, and overall linguistic difficulty align with the original problem $p$ and grade reading level $g$, using the method from \cite{norberg2025linguistic}.
    \item \textit{Solvability Agent}. The solvability validator agent solves the problem to verify mathematical consistency with the original and checks answer-set validity (i.e., the correct option is present in multiple-choice items).
    \item \textit{Authenticity Agent}. The authenticity validator agent checks whether the personalized context is age-appropriate and relatable to middle school learners.
\end{itemize}

\subsubsection{Refinement Agent}
The refinement agent uses feedback from validator agents and updates the candidate problem $\hat{p}$. This agent is instructed to preserve the original mathematical structure of the problem $p$, the target topic $t$, and the grade reading level $g$, while taking feedback $f_i,\,i\in\{1,2,3,4\}$ into account.

\subsection{Refinement Strategies}\label{method:rs}

\begin{table}[t]
\centering
\footnotesize
\renewcommand{\arraystretch}{1.15}
\setlength{\tabcolsep}{5pt}
\caption{Example outputs for \textit{Conversion Agent}, \textit{Validator Agents}, and \textit{Refinement Agent} given initial problem and topic.}
\begin{adjustbox}{max width=\linewidth}
\begin{tabularx}{\linewidth}{Y}
\toprule
\textbf{Problem Refinement Example (Topic: Basketball).} \\
\midrule

\textbf{Initial Problem:} The distance from the tip of a slice of pizza to the crust is 7 in.  Find diameter, radius, or circumference \\

\textbf{Step 1 (Conversion Agent):} Converting initial problem  \\

$\rightarrow$ \emph{\textbf{Candidate \#1:}}On a basketball court, the center circle is painted on the floor. The distance from the center dot to the painted line of the circle is 7 in. Find  diameter, radius, or circumference.\\

\textbf{Step 2 (Validator Agents):} \textit{fail}; \textit{feedback}: 7 inches would make the circle unrealistically small \\

\textbf{Step 3 (Refinement Agent):} Revising candidate given feedback \\

$\rightarrow$ \emph{\textbf{Candidate \#2:}} On a basketball court, the center circle is painted on the floor. The distance from the center dot to the painted line of the circle is 7 ft.Find  diameter, radius, or circumference.\\

\textbf{Step 4 (Validator Agent):} \textit{pass}. \\

\bottomrule
\end{tabularx}
\end{adjustbox}
\label{tab::example_refinement}
\end{table}

We now detail the three refinement strategies we explore to coordinate the validator and refinement agents. All strategies iterate for at most $k$ iterations. One refinement iteration consists of: (1) validating the current candidate problems with all validator agents and (2) applying strategy-specific revisions to produce the next candidate problem, utilizing one or more refinement agents. Refinement stops early if the candidate problem passes all validator agent checks.

\subsubsection{Centralized Refinement}
In Centralized Refinement, the feedback aggregator concatenates all validator agent feedback from those who failed their validations into a single message. A single refinement agent then generates a revised candidate conditioned on $(p,t)$ and the aggregate feedback $f_a \subset \{f_1, f_2, f_3, f_4\}$. This loop repeats until the candidate passes all validator agents or we have reached the maximum refinement iterations.

\subsubsection{Centralized Refinement with Planning}
In Centralized Refinement with Planning, a \textit{planning agent} is inserted between feedback aggregation and revision. The planning agent converts the unstructured aggregated validator feedback $f_a$ into a structured, prioritized action plan (e.g., a checklist of edits). The plan prioritizes correctness-preserving fixes (e.g., mathematical consistency and valid solution/choices) before stylistic refinements (e.g., wording and readability). The refinement agent then executes the plan to produce the next candidate problem. Table~\ref{tab:revision_plan} shows an example action plan.

\subsubsection{Decentralized Refinement}
In Decentralized Refinement, validation and revision are paired. When a specific validator agent $i$ fails, it triggers a corresponding specialized refinement agent that addresses only feedback from that agent, $f_i$. Revisions are applied sequentially in a fixed priority order. This setup reduces cascading errors and avoid information overload that can arise from aggregating all feedback at once \cite{wang-etal-2025-anymac}. We use the priority ordering of correctness $\rightarrow$ realism $\rightarrow$ authenticity $\rightarrow$ readability, established based on teacher feedback from prior work: mathematical correctness must be preserved before contextual or stylistic refinements, since such changes can inadvertently introduce mathematical errors \cite{Walkington2025,jia-etal-2025-eduagentqg}. Each pair of validator/refinement agent runs at most once per iteration; the updated candidate problem is then passed to the next validator.

\begin{table}[t]
\centering
\footnotesize
\renewcommand{\arraystretch}{1.03}
\setlength{\tabcolsep}{4pt}
\caption{Example plan generated from the \textit{Centralized with Planning} strategy to improve realism. The plan is generated with \textit{Conversion Agent} output, \textit{Candidate \#1}, from Table~\ref{tab::example_refinement}.}
\label{tab:revision_plan}

\begin{tabular}{@{}p{0.06\linewidth}@{\hspace{0.02\linewidth}}p{0.88\linewidth}@{}}
\hline
\textbf{\#} & \textbf{Plan (edits to improve the problem)} \\
\hline

\textbf{1} &
\textbf{Fix the realism/discreteness issue (highest priority).}\par
\textendash\ Change the given measurement to a realistic size for a basketball court center circle (use feet or a realistic number of inches).\par
\textendash\ Target a radius around 6 ft (72 in). \\[2pt]

\textbf{2} &
\textbf{Decide whether the context should stay ``basketball court'' or be changed.}\par
\textendash\ If keeping the basketball court setting, ensure all quantities and units match that scale (feet/yards, not small inches).\par
\textendash\ If you want to keep ``7 inches'' specifically, switch the object to a small circular basketball-related item. \\[2pt]

\textbf{3} &
\textbf{Ensure answer choices and follow-up tasks align with the corrected scale.}\par
\textendash\ Require a variable definition such as \(h=\) number of hours after noon, tied to the context. \\[2pt]

\textbf{4} &
\textbf{Add a light authenticity detail without adding complexity.}\par
\textendash\ Include a simple, realistic court-related detail. \\

\hline
\end{tabular}
\end{table}

\section{Evaluation}

In this section, we experimentally evaluate the effectiveness of our multi-agent framework, quantitatively compare the three refinement strategies, conduct a human evaluation on the validator agents, and show a qualitative case study.

\subsection{Quantitative Analysis}
\subsubsection{Dataset}
We experiment on 600 math problems drawn from an online mathematics homework platform, ASSISTments \cite{assistments}. Each problem is personalized to a topic drawn from a fixed set of 20 student interest topics collected by the platform. These topics span multiple categories, such as sports, entertainment, and media.

\subsubsection{Experimental Settings} 
We evaluate all strategies under identical generation and refinement settings to ensure a fair comparison. We set a maximum of $k{=}3$ refinement iterations for all strategies and use GPT-5.2 with temperature$ = 0$ to generate deterministic, and directly comparable outputs. In addition to the refinement strategies, we include a \textit{Zero-Shot} baseline, defined as the direct output of the conversion agent before any validator-driven feedback or revision.

\subsubsection{Quantitative Comparison across Refinement Strategies} 
\begin{figure}[tp]
  \centering
  \includegraphics[width=1\linewidth]{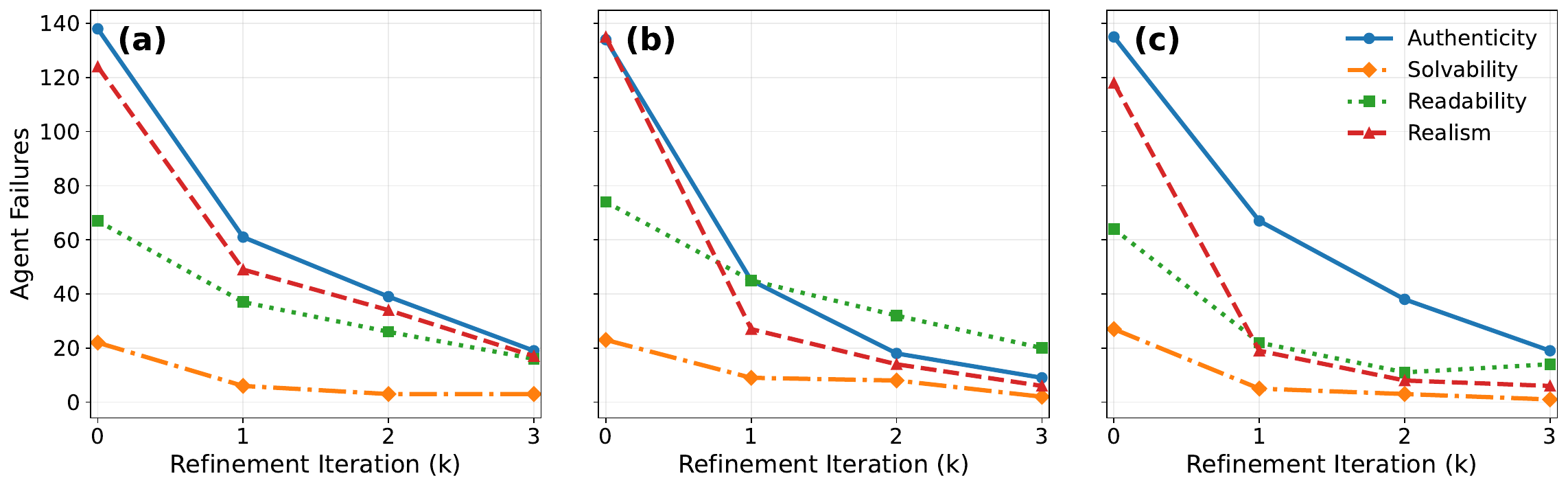}
  \caption{Validator agent failures across three refinement iterations, aggregated over 600 problems for three refinement strategies: (a) \textit{Centralized Refinement} (b) \textit{Centralized Refinement with Planning} (c) \textit{Decentralized Refinement.}}
  \label{fig:iters}
\end{figure}


Figure \ref{fig:iters} demonstrates validator agent failure counts across three refinement iterations for the three refinement strategies. 
We  observe that most improvements occur early on, with substantial reductions in failure counts after the first refinement iteration. This observation suggests that the multi-agent framework effectively identifies and corrects major feedback in the initial generation, with a smaller reduction in failure rates in subsequent iterations. 

For solvability, all strategies start with a low failure count and rapidly converges to near zero after a single iteration. This trend aligns with findings in prior work and suggests that recent LLMs are highly capable of producing mathematically valid problems. This result is comparable to human benchmarks, in which teachers report mathematical errors in approximately 10\% of cases \cite{christ-etal-2024-mathwell}.

For realism and readability, we observe that Decentralized Refinement converges to a low failure count faster than both Centralized strategies, achieving minimal failure counts by the second refinement iteration. One possible explanation is that in Decentralized Refinement, each validator has its own specialized refinement agent focusing exclusively on a single criterion, addressing specific feedback without needing to simultaneously satisfying other constraints. In contrast, Centralized and Centralized with Planning require revisions that balance multiple criteria simultaneously, which creates a harder task for the refinement agent. We postulate that for feedback that can be addressed with localized edits to a candidate problem, e.g., ones related to realism and readability, we should not aggregate this feedback with other feedback before refinement.

For authenticity, we observe that Centralized with Planning converges quicker and to the lowest failure count after the first refinement iteration. Decentralized Refinement, however, converges more slowly. One possible explanation is in the ordering of the refinement agents: since the readability refinement agent acts after the authenticity refinement agent, these revisions may undo fixes that promote authenticity. The better performance of Centralized strategies may also be attributed to the fact that authenticity is usually highly correlated with other criteria. Authenticity may also be more global and coherence-dependent than the other criteria. In contrast, realism, solvability, and readability can often be improved with localized edits, such as adjusting numerical values in the candidate problem, adding a missing statement, and simplifying wording, without changing the overall setup of the problem.

\subsubsection{Human Evaluation on Validator Agent Reliability}

\begin{table}[t]
\centering
\caption{ Fleiss’ $\kappa$ measuring inter-annotator agreement across the three annotators. Accuracy and Cohen’s $\kappa$ between the human majority label (Maj) and the model output.}
\renewcommand{\arraystretch}{1.15}
\setlength{\tabcolsep}{6pt}
\resizebox{\linewidth}{!}{%
\begin{tabular}{l r r r r}
\hline
\textbf{Criterion} & \textbf{n} & \textbf{Fleiss' $\kappa$ (3 ann.)} & \textbf{Acc. (Maj. vs Model)} & \textbf{Cohen's $\kappa$ (Maj. vs Model)} \\
\hline
Realism      & 45 & 0.310 & 0.667 & 0.322 \\
Authenticity & 45 & 0.023 & 0.511 & 0.133 \\
Readability  & 45 & 0.095 & 0.800 & 0.281 \\
\hline
\end{tabular}%
}
\label{tab:agreement_kappas}
\end{table}

To assess validator reliability, we conduct a human evaluation. We recruit three annotators with a college-level mathematics background to evaluate 45 problems sampled from our \textit{Zero-Shot} baseline outputs. These problems include ones that both passed and failed the validator agent checks. Annotators assigned pass/fail labels for authenticity, realism, and readability using the same rubric as LLM-based validator agents described in Section \ref{sec:val_agents}. We excluded solvability since its labels were highly skewed (few failures), making agreement estimates less informative.

In Table~\ref{tab:agreement_kappas}, we report Cohen's $\kappa$ \cite{cohen1960coefficient} and accuracy (i.e., percentage of time the labels match) between the human majority label and the validator agent output. The human majority label refers to the most frequently selected label (pass or fail) across the three annotators, where no ties are possible since labels are binary. 

For realism ($\kappa{=}0.322$, $\text{acc}{=}0.667$) and readability ($\kappa{=}0.281$, $\text{acc}{=}0.800$), there is \emph{fair} agreement between validator agents and human annotators. For authenticity ($\kappa{=}0.133$, $\text{acc}{=}0.511$), there is only \emph{slight} agreement. We also calculate human inter-annotator agreement using Fleiss' $\kappa$ \cite{fleiss1971measuring} across three annotators. For realism (Fleiss' $\kappa{=}0.310$), there is \emph{fair} agreement. For authenticity (Fleiss' $\kappa{=}0.023$) and readability (Fleiss' $\kappa{=}0.095$), there is \emph{slight} agreement. Therefore, the agreement between humans and LLM-based validators are roughly in the same range.

For readability, accuracy ($\text{acc}{=}0.800$) indicates that the validator agent often matches the majority decision, even when annotators disagree. The low $\kappa{=}0.281$ likely stems from the fact that readability failures are rare, since most problems pass the readability assessment. 
For realism, the LLM-based validator agent has fair agreement with human judgment, comparable to inter-annotator agreement. This result suggests the validator captures a reasonably consistent notion of realism that aligns with human judgment. 

For authenticity, the validator shows limited reliability, with a near-random accuracy level ($\text{acc}{=}0.511$) and agreement barely above zero. This low reliability appears to stem from genuine subjectivity in authenticity judgments and differences. Moreover, there can be significant differences in domain-specific knowledge between the validator and human annotators, particularly for contexts grounded in pop culture or specialized practices. For example, the problem statement ``At baseball practice, Jordan throws 7.1 innings in one game and 3.4 innings in another game...'' was assigned a label of fail for realism and authenticity by the validator agents, because innings are conventionally expressed in thirds (e.g., $7\frac{1}{3}$), not in tenths. In contrast, our human annotators, likely without knowledge on baseball, labeled this example as pass. Therefore, while validator agents can reliably assess more objective criteria such as realism and readability, they may require domain-specific knowledge or human-in-the-loop verification for subjective criteria such as authenticity.

\begin{table}[t]
\centering
\fontsize{8.5}{10}\selectfont
\setlength{\tabcolsep}{6pt}
\renewcommand{\arraystretch}{1.08}
\caption{Top-3 topics and curriculum units showing the highest prevalence of each validator failure type in \textit{Zero-Shot} outputs.}
\label{tab:topk_failure_types_topic_unit}
\begin{tabularx}{\linewidth}{l p{0.25\linewidth} X}
\toprule
\textbf{Failure Type} & \textbf{Topic} & \textbf{Curriculum Unit} \\
\midrule
\multirow{3}{*}{Authenticity}
 & Classical Music & Putting it All Together \\
 & Twitter         & Angles, Triangles, and Prisms \\
 & TikTok          & Probability and Sampling \\
\midrule
\multirow{3}{*}{Realism}
 & Baseball   & Rational Numbers Arithmetic \\
 & Mario Kart & Measuring Circles \\
 & NY Knicks  & Introducing Proportional Relationships \\
\midrule
\multirow{3}{*}{Readability}
 & Arctic Monkeys & Scale Drawings \\
 & NY Knicks      & Expressions, Equations, and Inequalities \\
 & Boston Red Sox & Measuring Circles \\
\midrule
\multirow{3}{*}{Solvability}
 & Roblox     & Proportional Relationships and Percentages \\
 & Josh Allen & Rational Numbers Arithmetic \\
 & Fast Food  & Probability and Sampling \\
\bottomrule
\end{tabularx}
\end{table}

To understand which validator failures occur in different contexts, we analyze failure patterns in \textit{Zero-Shot} outputs across topics and math curriculum units. Math curriculum units are the overarching math content areas that a math problem is associated with. For each failed problem, we record which validator criteria failed and compute failure prevalence within each topic and curriculum unit. Prevalence is computed as the proportion of failed attempts within a given topic or curriculum unit with a failure type.

Table~\ref{tab:topk_failure_types_topic_unit}  presents the top topics for each failure type, showing which failures are most prevalent within different topics. We find that failure patterns vary by topic. Topics related to media (e.g., Twitter, TikTok) are linked to authenticity failures, showing that surface-level topical alignment does not ensure age-appropriate or relatable contexts. Topics related to sports and games are more often linked to realism and readability failures, suggesting difficulty in keeping quantities plausible and language accessible while preserving the original math. Solvability failures are generally infrequent, with no clear pattern across topics.

Table~\ref{tab:topk_failure_types_topic_unit}  also presents the top curriculum units for each failure type, showing which failures are most prevalent within different units. Curriculum units involving measurement or proportions are associated with realism failures, indicating difficulty keeping quantities plausible during personalization. Curriculum units that involve direct calculations, such as solving expressions, are associated with readability failures, since rewriting can increase sentence length or vocabulary difficulty. Curriculum units associated with authenticity failures span a variety of subjects, indicating that these issues are widespread rather than domain-specific. These results highlight the complexity of the context personalization task, as different failure types are prevalent in different topics and units.

\subsection{Case Analysis}


\begin{table}[t]
\caption{Refinement outputs from three distinct refinement strategies.}
\centering
\small
\setlength{\tabcolsep}{4pt}
\renewcommand{\arraystretch}{1.15}
\begin{tabularx}{\linewidth}{@{}>{\raggedright\arraybackslash}p{2.2cm}>{\raggedright\arraybackslash}X@{}}
\toprule
\textbf{Version} & \textbf{Problem Text} \\
\midrule
Original &
Find the product. $(-7)\cdot(-1.1)$ \\
\addlinespace[2pt]
Centralized &
A Led Zeppelin fan writes down two numbers about the band: $-7$ and $-1.1$. What number do you get when you multiply $-7$ by $-1.1$? \\
\addlinespace[2pt]
Centralized with Plan &
A fan is buying Led Zeppelin stickers. Each sticker costs \$1.10. The fan buys 7 stickers. How much money does the fan spend in all? \\
\addlinespace[2pt]
Decentralized &
In a Led Zeppelin trivia game, the score change is represented by $(-7)\cdot(-1.1)$. What is the product of $(-7)$ and $(-1.1)$? \\
\bottomrule
\end{tabularx} \label{tab:prabmsh4_examples}
\end{table}

\textbf{Analysis of Strategy Trade-offs} To illustrate how the three refinement strategies balance competing constraints, we  selected a problem where all three strategies succeed at different criteria but make distinct trade-offs. Table~\ref{tab:prabmsh4_examples} shows refined versions of a signed multiplication problem personalized to ``Led Zeppelin." This topic-problem pair illustrates a common personalization challenge: Led Zeppelin resonates with students interested in classic rock, but creating an authentic Led Zeppelin context that naturally involves multiplying two negative numbers requires balancing of mathematical structure and realistic scenarios.

Centralized Refinement largely preserves the original mathematical form, retaining the signed operands verbatim while changing the topic superficially, where a Led Zeppelin fan multiplies two numbers. This strategy maintains equation-level faithfulness to the original problem but provides limited context for the quantities involved, leading to reduced realism and authenticity. 

Centralized with Planning improves contextual realism by mapping the computation onto a familiar real-world schema. This reframing creates a more concrete and relatable context. However, this transformation changes the original mathematical form by converting negative number multiplication into positive number multiplication. If the teacher's objective is specifically to practice multiplying negative values, this modification would not be pedagogically appropriate.

Decentralized Refinement aligns narrative elements more explicitly with the original mathematical form, by framing the computation as point calculation in a trivia game, leading to better coherence between the topic and the problem's mathematical structure. However, it introduces a logical inconsistency: ``losing 1.1 points for 7 rounds'' naturally corresponds to $7 \cdot (-1.1)$ rather than $(-7) \cdot (-1.1)$, weakening the alignment between the story and the underlying mathematical structure. 

These differences illustrate the trade-offs in each refinement strategy. Centralized emphasizes mathematical preservation at the expense of contextual richness, Centralized with Planning emphasizes realism but may compromise pedagogical intent, while Decentralized emphasizes narrative coherence but increases sentence complexity. These trade-offs underscore the inherent difficulty of context personalization when constrained to specific topics, where authentic real-world scenarios may not naturally align with the structure of the original problem.

\section{Conclusions and Future Work}
This work investigates how to make LLM-based \emph{context personalization} of math word problems usable in practice by explicitly validating and refining across four teacher-motivated criteria: solvability (mathematical consistency), realism (plausible quantities and contexts), readability (grade-appropriate language), and authenticity (age-appropriate, relatable contexts). Across our experiments we find two consistent patterns. First, the most common initial weaknesses of personalized problems are \emph{authenticity} and \emph{realism}. Second, most quality gains occur early, often after a single generate--validate--revise iteration. Additionally, our experiments demonstrate that the differences among the refinement strategies are criterion dependent.

There are many avenues for future work. First, we can improve our agent design and coordination using more sophisticated strategies for adaptive agent orchestration. For example, if authenticity benefits from holistic changes but solvability benefits from direct revision, we may coordinate the corresponding agents in different ways. Second, our task may also benefit from topic and lesson specific constraints. In the case where we are personalizing mathematical expression, we may need to loosen our readability constraints as the problem will increase in reading complexity. Finally, we need to conduct larger-scale human evaluation, preferably including teachers and actual students, to measure validator reliability more accurately, especially for authenticity, and ultimately study the impact of math problem personalization on learning outcomes.

\subsubsection{Acknowledgments} This work is partially supported by Renaissance Philanthropy under the learning engineering virtual institute (LEVI) initiative and the NSF under grants 2237676 and 2341948. 


\bibliographystyle{splncs04}
\bibliography{main}

\end{document}